\documentclass[twocolumn,showpacs,preprintnumbers,amsmath,amssymb]{revtex4}
\usepackage{graphicx}% Include figure files
\usepackage{dcolumn}% Align table columns on decimal point
\usepackage{bm}% bold math

\begin{document}

\preprint{APS/123-QED}

\title{Direct measure of the exciton formation in quantum wells from
time resolved interband luminescence}% Force line breaks with \\
\author{J.~Szczytko, L.~Kappei, J.~Berney, F.~Morier-Genoud, M.T.~Portella-Oberli,
B.~Deveaud}

\affiliation{Institut de Photonique et Electronique Quantiques,
Ecole Polytechnique F\'{e}d\'{e}rale de Lausanne (EPFL) CH1015
Lausanne, Switzerland}

\date{\today}% It is always \today, today,
             %  but any date may be explicitly specified

\begin{abstract}
We present the results of a detailed time resolved luminescence
study carried out on a very high quality InGaAs quantum well
sample where the contributions at the energy of the exciton and
at the band edge can be clearly separated. We perform this
experiment with a spectral resolution and a sensitivity of the
set-up allowing to keep the observation of these two separate
contributions over a broad range of times and densities. This
allows us to directly evidence the exciton formation time, which
depends on the density as expected from theory. We also evidence
the dominant contribution of a minority of excitons to the
luminescence signal, and the absence of thermodynamical
equilibrium at low densities.
\end{abstract}


\maketitle

%\section{\label{Intro}Introduction}
Excitons in quantum wells form quite an appealing quasiparticle
showing a large range of optical properties that have proven at
the same time technologically useful, and physically interesting
\cite{Chemla1982}. A large part of this interest is linked with
the appearance of excitonic resonances in absorption up to room
temperature. It is also well known, since the seminal work of
Weisbuch et al \cite{Weisbuch1981}, that free excitons appear to
dominate the luminescence response of semiconductor quantum wells
at low temperatures. Part of the origin of this effect lies in the
breakdown of the translational symmetry which brings a very
efficient recombination channel to free excitons in quantum wells
\cite{Agranovitch1966,Deveaud1991}.

Interestingly, in the low density regime, time resolved
luminescence (TR-PL) in quantum wells is observed to be always
dominated by light coming at the exciton energy, even under non
resonant excitation. The observations of this dominant
contribution are so numerous that only a very partial list of
references may be given here
\cite{Deveaud1987,Yoon1996,Damen1990,Robart1995,Deveaud1993} (in
order to be specific, we only consider here the case of quantum
wells grown on GaAs substrates). A double question has then been
debated for more than 10 years in the literature: first how do
free electron hole pairs bind into excitons and second does indeed
the luminescence at very short time proceed from bound excitons. A
brief survey of the literature allows to find that
experimentalists have reported formation times ranging from less
than 10 ps up to about 1 ns
\cite{Damen1990,Robart1995,Deveaud1993,Kaindl2003} and
theoretical values range from 100 ps
\cite{Thilagam1993,Piermarocchi1996,Piermarocchi1997} to over 20
ns \cite{Hoyer2003}. Clearly, the origin of this spreading in the
reported values lies in the poor sensitivity of the experiments
used in general to probe the exciton formation process, except for
the case of the recent terahertz absorption experiments
\cite{Kaindl2003}. On the theoretical side, binding of an
electron hole pair into an exciton requires, at low temperatures,
the emission of an acoustic phonon, which brings long formation
time due to the small coupling of acoustic phonons to excitons.

The long formation time of excitons, together with the
observation of luminescence at the exciton energy at the shortest
times \cite{Damen1990,Hayes2002} led Kira et al \cite{Kira} to
introduce the idea that a free electron hole plasma, properly
including Coulomb correlation effects, should give rise to
luminescence at the exciton energy, without any exciton
population. Although this proposed interpretation is currently
largely questioned (see for example Hannewald and Glutsch et al
\cite{Hannewald2000}), the dominance of the excitonic transition
at the shortest time has not received a sensible explanation yet.

In the present work, we use a properly designed sample with a
particularly high quality, together with a luminescence set-up
with improved sensitivity, to study the exciton formation and to
give clues on the origin of luminescence in quantum wells. We
will show that the formation times of excitons depend on the
excitation density as expected from theory. We will also evidence
how a very small population of excitons has a large enough
recombination rate to dominate the luminescence spectrum, even at
the shortest times accessible in the present experiment.

We have selected a particular sample, because of its unequalled
quality. It consists of a single In$_{x}$Ga$_{1-x}$As  80 \AA\
quantum well (QW), with a low indium content of about $x=5\%$
grown by molecular-beam epitaxy. This QW is embedded in the
middle of a GaAs layer of total mean thickness $\lambda$ (where
mean $\lambda$ corresponds to the wavelength of the excitonic
resonance in the QW), which was grown over a 10 period
distributed Bragg reflector (DBR). This DBR allows to measure the
absorption of the sample in the reflection configuration without
any sample preparation. It also increases the optical coupling of
the QW, but do not disturb the shape of the observed luminescence
spectrum, because the resonance mode has a spectral width of
about 40~nm. Such a DBR changes slightly the radiative properties
of free carriers, but does not affect their relaxation properties
which we are studying here. The spectrum was recorded with a CCD
camera in cw and with a streak camera in the time-resolved
experiment (resolution of 3~ps, photon-counting mode). The
temporal resolution of the whole setup is limited to about
20--30~ps, because of the spectral resolution of 0.1 meV. The
excitation energy we use was $\hbar\omega$ = 1.5174~eV with the
power within the ranges 1.0--300~$\mu$W (photon density
$N_{\nu}=9\times 10^8$--$3\times 10^{11}$ photons/cm$^{2}$ per
pulse; 85~$\mu$m spot). More details about the sample and
experimental setup can be found in \cite{Szczytko2003}.

\begin{figure}
  \includegraphics[width=0.9\columnwidth]{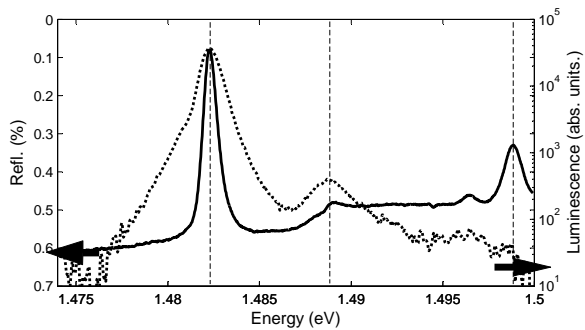}
\caption{\label{fig1} Cw absorption (calculated as
$1-\text{Reflectivity}$; bold line, left axis) and the TR-PL
integrated over 1300 ps of the sample (logarithmic scale, right,
for luminescence). The structure at 1.4823 eV corresponds to the
$1s$ heavy-hole exciton, at 1.4888 eV to the plasma transition
and at 1.4988 to the light-hole exciton (vertical lines). The low
energy exponential tail of the excitonic transition originates
from the trion transition at 1.4807 eV (discussed in
\cite{Szczytko2003}). }
\end{figure}

The high quality of the sample is evidenced through optical
measurements. We do not observe any Stokes shift between the
absorption and luminescence at $E_{\text{G}}$ and $E_{1s}$
(Fig.~\ref{fig1}). Moreover the observed lifetime of the QW
reaches 3.7~ns in the low density limit at 10K, which indicates
the very low density of non-radiative recombination centers. The
linewidth of the exciton is less than 1.0~meV and mainly given by
the homogeneous broadening as the lorentzian lineshape shows. The
luminescence from free carriers appears exactly at the position
of the band gap, which is known from the measurement of the
absorption of the sample at the same position (Fig.~\ref{fig1}).
The free carrier luminescence shows the expected high energy
exponential tail, corresponding to the Boltzmann distribution of
the carriers, with a temperature given by the temperature of the
lattice for temperatures above 25 K \cite{Szczytko2003}.

\begin{figure}
  \includegraphics[width=0.8\columnwidth]{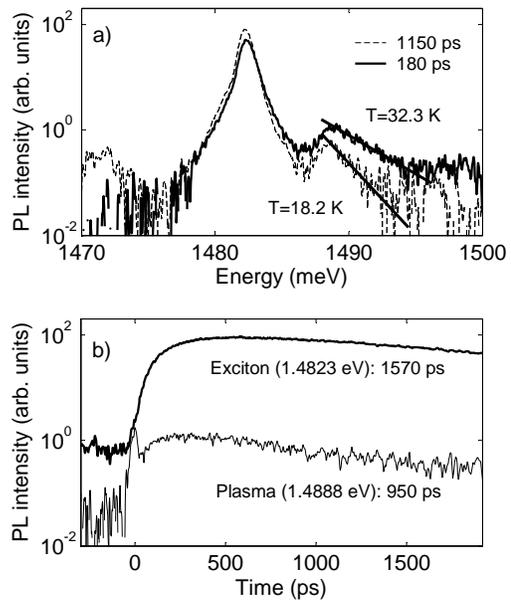}
\caption{\label{fig2} Spectral a) and temporal b) traces of the
TR-PL. $N_{\nu}=2\times$10$^{10}$ photons/cm$^{2}$. a) the
temperatures of the carriers are estimated 180 ps and 1150 ps
after the excitation. b) the temporal dependence of the maxima of
the excitonic and plasma transition. The small peak seen at $t=0$
for plasma transition is caused by the diffused light of the
laser pulse.}
\end{figure}

\begin{figure}
  \includegraphics[width=0.9\columnwidth]{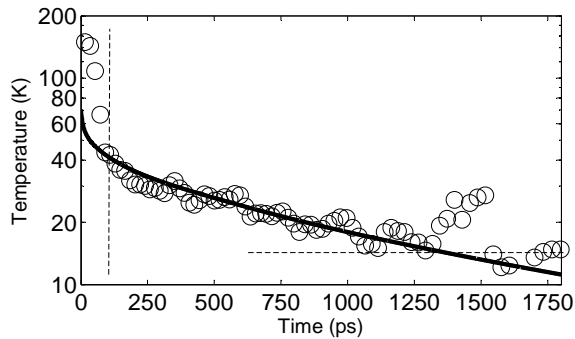}
\caption{\label{fig3} Temperature of the carriers deduced from
the exponential slope of the luminescence above gap
($N_{\nu}=2\times$10$^{10}$ cm$^{-2}$). The full line corresponds
to the theoretical cooling curve.}
\end{figure}

The main interest of the sample is that the transitions at above
gap energies can be resolved at all times and densities. Typical
time-resolved luminescence traces are shown in Fig.~\ref{fig2}
for the spectral and temporal domain. The sufficient dynamical
range allows us to measure the free carrier temperature directly
from the spectrum for all times after 100 ps delay. Although it
is possible to fit the exponential decay to the slope of the
plasma transition for $t<100$ ps (these results are also shown in
Fig.~\ref{fig3}) we think that these results might not correspond
to the real temperature. Likewise due to the finite linewidth of
the free carrier transition there is the finite minimal
temperature which we can determine. This reduces the time
interval in which we can attribute the temperature precisely to
about 100--1300 ps.

The thermalization of hot carriers can be modeled by computing the
average energy-loss rate per electron-hole pair  $\langle d E /d
t\rangle \propto \langle d T /d  t\rangle$  and deducing the
variations of their temperature from:
\begin{equation}\label{eloss}
 T(t)=T_0 - \int_0^t \langle \frac{d T }{d  t}\rangle dt
\end{equation}
where $T_0$ is the carrier temperature at $t=0$. Knowing the
theoretical curve $\langle d E /d t\rangle$ for a Boltzman
distribution in the case of bulk GaAs \cite{Shah1984,Leo1988} one
can easily calculate the cooling curve by numerical integration
with the only parameter $T_0$. However in the case of QWs an ad
hoc additional factor $\alpha > 1$ is added $\langle d E /d
t\rangle_{\text{exper}} = \langle d E /d t\rangle_{\text{theor}}
/\alpha$ by which the measured energy-loss rate is reduced
compared to the theoretical value
\cite{Shah1984,Leo1988,Ruhle1989}. Here we obtain a good fit for
$\alpha =2.9$, see the bold line in Fig.~\ref{fig3}.

\begin{figure}
  \includegraphics[width=0.9\columnwidth]{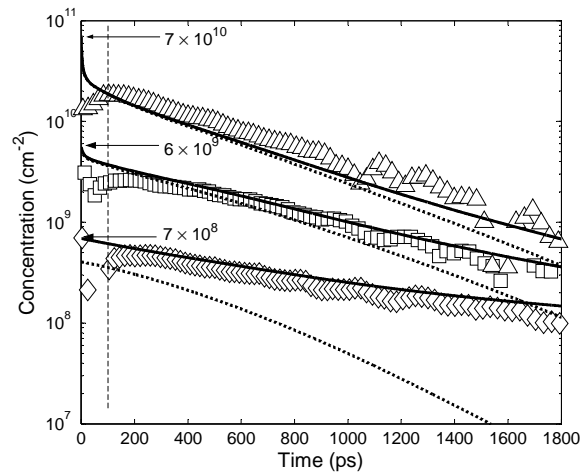}
\caption{\label{fig4} Concentration of free carriers as a functon
of time. The experimental data calculated according Eq.~\ref{nnn}
for three different absorbed photon densities are shown together
with the time decay fitting curves (solid lines,
Eq.~\ref{ratioeqnNX}). The corresponding $N_{\text{eq}}$
calculated with Eq.~\ref{saha} are shown as a dotted lines. The
number of absorbed photons is given per pulse per cm$^2$. The
vertical line is set for the time of 100 ps. }
\end{figure}

Very importantly, the shape of the luminescence signal above the
gap is the same within the two main theoretical descriptions
(excitons plus free carriers \cite{Ridley1990,Yoon1996}, or
Coulomb correlated free carriers \cite{Kira}). In both cases, we
expect a Boltzmann like luminescence line provided electrons and
holes are thermalized, which is the case for times longer than
100 ps. As will be shown below the luminescence of free carriers
above the band edge provides a direct measure of the relative
variations of the population of free electrons and holes. Indeed,
in both models, the intensity at each energy is simply
proportional to the product of the associated distribution
functions $f_e$ and $f_h$. Then, in the case of the low density
regime we are studying, the integrated intensity of the free
carrier luminescence is proportional to the concentration of the
electrons $n$ and holes $p$ via the bimolecular recombination
rate $B$ \cite{Lasher1964}.
\begin{equation}\label{brr}
  I_{\text{plasma}}\sim B n p = B n^2
\end{equation}
as we have $n=p$. The parameter $B$ at low temperatures in QW is
simply inversely proportional to the temperature $T$ of the
carriers  \cite{Matsusue1987}:
\begin{equation}\label{btheoret}
  B=\frac{8 \pi e^2 n_r}{m_0^2 c^3 (m_c+m_v)} \langle  \left|P_{cv}\right|^2 \rangle
  \frac{E_{g}}{k_BT}
\end{equation}
where $n_r$ is the refractive index, $c$ is the light velocity,
$m_0$ is the free-electron mass, $m_c$ and $m_v$ are the reduced
masses of the electron and holes, $\langle  \left|P_{cv}\right|^2
\rangle$ is the squared momentum matrix element between electron
and hole in each subband averaged over directions and
polarizations of photons and over spins of electron end holes,
$E_{g}$ is the energy gap in QW \cite{Matsusue1987}.
Eq.~\ref{brr} and \ref{btheoret} lead to
\begin{equation}\label{nnn}
  n\propto \sqrt{I_{\text{plasma}} T}
\end{equation}
Thus knowing the time evolution of both quantities -- the free
carriers luminescence intensity $I_{\text{plasma}}(t)$ and their
temperature $T(t)$ (Eq.~\ref{eloss}) -- we can deduce in a very
simple and direct way the temporal evolution of the photoexcited
free carrier density. This evolution is plotted in
Fig.~\ref{fig4} (symbols) and is obtained from the results shown
in Fig.~\ref{fig2} and \ref{fig3} \footnote{to calculate
$I_{\text{plasma}}$ we integrate the intensity of the free
carrier transition in the range 1.487-1.495 eV.}.

This temporal evolution is governed by four factors: the
electron-hole radiative recombination rate $B$, the non-radiative
decay time $\tau_{\text{nr}}$  and the formation and ionization of
the excitons -- via the bimolecular formation coefficient $C$. We
get two rate equations for the population of free carriers $n$
and of excitons $X$:
\begin{subequations}\label{ratioeqnNX}
\begin{eqnarray}
  \frac{d\;n}{d\;t}&=&  - \gamma C n^2 +  \gamma C N_{\text{eq}}^2
  -\frac{n}{\tau_{\text{nr}}} - B n^2
\label{ratioeqn}
  \\
  \frac{d\;X}{d\;t}&=& \gamma C  n^2 -  \gamma C N_{\text{eq}}^2 - \frac{X}{\tau_{\text{D}}}\label{ratioeqnX}
\end{eqnarray}
\end{subequations}
%\begin{subequations}\label{ratioeqnNX}
%\begin{eqnarray}
%  {d\;n}/{d\;t}&=&  - C n^2 +  C N_{\text{eq}}^2
%  -{n}/{\tau_{\text{nr}}} - B n^2
%\label{ratioeqn}
%  \\
%  {d\;X}/{d\;t}&=& C  n^2 -  C N_{\text{eq}}^2 - {X}/{\tau_{\text{D}}}\label{ratioeqnX}
%\end{eqnarray}
%\end{subequations}
where $C$ is the rate calculated by Piermarocchi et al
\cite{Piermarocchi1997}. It depends upon both carrier and lattice
temperature through the interaction with optical and acoustic
phonons. $\gamma$ is a multiplication factor by which the measured
formation rate in our InGaAs QW is changed compared to the
theoretical value obtained for GaAs. $N_{\text{eq}}(n,X,T)$ is
the equilibrium carrier concentration given by the Saha equation
(for $n=p=N_{\text{eq}}$) \cite{Philips1996}:
\begin{equation}\label{saha}
    \frac{N_{\text{eq}}^2}{X}=K(T)=\frac{\mu_{X}k_B T}{2 \pi
    \hbar^2}\exp (\frac{-E_b}{k_B T})
\end{equation}
%\begin{equation}\label{saha}
%    {N_{\text{eq}}^2}/{X}=K(T)={\mu_{X}k_B T}/{\pi
%    \hbar^2}\exp ({-E_b}/{k_B T})
%\end{equation}
where $\mu_X$ is the reduced mass of the exciton and $E_b$ is the
exciton binding energy. The therm $C N_{\text{eq}}^2$, in the
case of cw excitation, leads to thermodynamical equilibrium
between the population of the excitons and carriers at the given
temperature. In time-resolved experiments this term drives the
system towards equilibrium.

Having a sample of unprecedented quality with lifetimes in excess
of 3 ns, we can safely neglect the non-radiative processes
$\tau_{\text{nr}}$. The radiative decay of the electron-hole
plasma is governed by the bimolecular processes. The initial
concentration is given by the absorbed photon density which can be
estimated from the excitation photon density $N_{\nu}$ and the
value of the absorption coefficient (i.e. Fig.~\ref{fig1}). In
our case the InGaAs QW absorbs about 30 \% of incoming photons.
Using the value of $B$ measured and computed by Matsutsue et al
$B = 10^{-3}$ cm$^2$/s \cite{Matsusue1987} we get typical
radiative decay times of the order of 20 ns for the temperatures
and densities of our experiment \cite{Philips1996}. Thus the
dominant term in Eq.~\ref{ratioeqn} comes from the formation of
excitons.

In Fig.~\ref{fig4} the results of our model calculations are
shown as solid lines. We used the values of the initial carrier
concentration estimated from the number of absorbed photons as
above. The time dependence of the temperature was taken from the
experiment (Fig.~\ref{fig3}). We took a constant value for the
thermalized exciton decay time $\tau_{\text{D}}=500$ ps (this
only weakly affects the fit). The only fitting parameter was
$\gamma=5$.

Although very simple, this model includes most important
ingredients of what should be a full theory, and includes the
results of the model by Piermarocchi
\cite{Piermarocchi1996,Piermarocchi1997}. Very importantly,
increasing the carrier density results in a faster formation of
excitons, because it leads to an increased probability of binding
one electron and one hole through interaction with an acoustic
phonon. This model also includes an initial drop of the carrier
concentration by the rapid formation of excitons through emission
of LO phonon by the initially hot carriers
\cite{Piermarocchi1997}. This process implies that, after
non-resonant excitation, a population of excitons of a few
percent may be formed within the first ps.

The parameters governing exciton formation are the equilibrium
density, depending on the temperature and density according to the
Saha equation and shown as dotted lines in Fig.4, and the
formation rate  $\gamma Cn$. The observed exponential decay of the
free carrier density with slopes from 500 ps to 1050 ps only
correspond directly to the exciton formation at the lowest
density, for the largest densities, the equilibrium is reached
within our temporal resolution, and the decay of the free carrier
population follows the changes imposed by the Saha equation. The
formation rate 1 ns after the excitation that fits best our
results is $1/60$ ps$^{-1}$ for the highest density used here and
decays to $1/1100$ ps$^{-1}$ for the lowest density, in very good
agreement with the recent time-resolved terahertz absorption
measurements \cite{Kaindl2003}.

It should be emphasized again that we do not assume
thermodynamical equilibrium between free carriers and excitons to
obtain this result. In fact for the lower excitation densities
the process of the exciton formation is so slow compared to the
radiative decay of excitons, that the system is never in
thermodynamical equilibrium! What we demonstrate here is that,
even for delays of the order of 200 ps, the population of
excitons is less than 10 \% of the total population at low
densities. However, a simple estimation of the rate of photon
emission through excitonic radiative emission shows that this
rate is roughly two orders of magnitude larger than the rate of
radiative recombination of free carriers, in good agreement with
our experimental observations. Therefore the excitonic transition
dominates the luminescence spectrum. We cannot be conclusive yet
about the relative importance of free carriers and excitons for
times shorter than 100 ps, and further experiments with an
improved sensitivity will have to be performed to answer this
point.

In summary, we have described here the results of the
time-resolved photoluminescence study of a very high quality
InGaAs quantum well sample. It has been possible to perform this
experiment with a spectral resolution and a sensitivity allowing
to keep track of the separate excitonic and free carrier
contribution over the whole time and density range. Thus we have
measured directly the temperature of the carriers and are able to
deduce the formation time of excitons. This time is measured over
two orders of magnitude in density, it changes from 30 ps for
densities of $10^{10}$ cm$^{-2}$ to 1000 ps for densities of
$3\times10^{8}$ cm$^{-2}$. We show that, even a very small
population of excitons may dominate the emission spectrum, at
least for delays longer than 100 ps.

Acknowledgements: this work was partially supported by the Swiss
National Research fund. We wish to thank T.~Guillet,
L.~Schifferle, J.-D.~Ganiere, S.~Koch, V.~Savona, D.~Chemla and
R.~Zimmermann for fruitful discussions.

\end{document}